\documentclass[aps,a4paper,floatfix,nofootinbib]{revtex4}

\usepackage[utf8]{inputenc}
\usepackage{graphicx,color}
\usepackage{amsmath,amssymb,amsfonts,amsthm,amscd,bm}
\usepackage{booktabs}
\usepackage{cancel}

\def\tilde{\widetilde}
\def\bar{\overline}
\def\hat{\widehat}
\def\*{\star}
\def\[{\left[}
\def\]{\right]}
\def\({\left(}      

\def\){\right)}

\def\zbar{{\bar{z} }}
\def\frac#1#2{\dfrac{#1}{#2}}
\def\inv#1{\dfrac{1}{#1}}
\def\half{\tfrac{1}{2}}
\def\d{\partial}

\def\2pi{\hbox{$2\pi i$}}

\def\dsl{\raise.15ex\hbox{/}\kern-.57em\partial}
\def\Dsl{\,\raise.15ex\hbox{/}\mkern-.13.5mu D}
   \def\CE{{\cal E}}

      \def\CO{{\cal O}}
      
\def\CS{{\cal S}}

\def\2pi{\hbox{$2\pi i$}}

\def\dsl{\raise.15ex\hbox{/}\kern-.57em\partial}
\def\Dsl{\,\raise.15ex\hbox{/}\mkern-.13.5mu D}
\font\numbers=cmss12
\font\upright=cmu10 scaled\magstep1
\def\stroke{\vrule height8pt width0.4pt depth-0.1pt}
\def\topfleck{\vrule height8pt width0.5pt depth-5.9pt}
\def\botfleck{\vrule height2pt width0.5pt depth0.1pt}
\def\Zmath{\vcenter{\hbox{\numbers\rlap{\rlap{Z}\kern
    0.8pt\topfleck}\kern 2.2pt
    \rlap Z\kern 6pt\botfleck\kern 1pt}}}
\def\Qmath{
    \vcenter{\hbox{\upright\rlap{\rlap{Q}\kern3.8pt\stroke}\phantom{Q}}}}
\def\Nmath{\vcenter{\hbox{\upright\rlap{I}\kern 1.7pt N}}}
\def\Cmath{\vcenter{\hbox{\upright\rlap{\rlap{C}\kern
                   3.8pt\stroke}\phantom{C}}}}
\def\Rmath{\vcenter{\hbox{\upright\rlap{I}\kern 1.7pt R}}}
\def\Z{\ifmmode\Zmath\else$\Zmath$\fi}
\def\Q{\ifmmode\Qmath\else$\Qmath$\fi}
\def\N{\ifmmode\Nmath\else$\Nmath$\fi}
\def\C{\ifmmode\Cmath\else$\Cmath$\fi}
\def\R{\ifmmode\Rmath\else$\Rmath$\fi}
\def\barray{\begin{eqnarray}}
\def\earray{\end{eqnarray}}
\def\beq{\begin{equation}}
\def\eeq{\end{equation}}

\def\n{\noindent}

\def\AA{\leavevmode\setbox0=\hbox{h}
\dimen0=\ht0 \advance\dimen0 by-1ex\rlap{\raise.67\dimen0\hbox{\char'27}}A}

\makeatletter
\def\iddots{\mathinner{\mkern1mu\raise\p@
\vbox{\kern7\p@\hbox{.}}\mkern2mu
\raise4\p@\hbox{.}\mkern2mu\raise7\p@\hbox{.}\mkern1mu}}
\makeatother

\theoremstyle{plain}

\theoremstyle{remark}

\def\UVtoIR{{\overset{UV\to IR}{\longrightarrow}}}

\def\half{\tfrac{1}{2}}

\def\psibar{\bar{\psi}}

\def\Q{Q_\infty}
\def\dim#1{[[ #1 ]]}
\def\mphys{m_{\rm phys}}
\def\gIR{{g_2^{\rm IR}}}
\def\gUV{{g_2^{\rm UV}}}
\def\s2{\sqrt{2}}
\def\adual{\tilde{a}}
\def\bdual{\tilde{b}}
\def\II{\mathcal{I}}

\begin{document}

\title{The sinh-Gordon model beyond the self dual point\\ and  the freezing transition in disordered systems}
\author{
Denis Bernard\footnote{denis.bernard@ens.fr} and  Andr\'e  LeClair\footnote{andre.leclair@gmail.com} 
}
\affiliation{Laboratoire de Physique de l'Ecole Normale Sup\'erieure, CNRS, ENS \& Universit\'e PSL, Sorbonne Universit\'e, Universit\'e de Paris, 75005 Paris, France}
\affiliation{Cornell University, Physics Department, Ithaca, NY 14850, USA} 

\medskip

\begin{abstract}
The S-matrix of the well-studied sinh-Gordon model possesses a remarkable strong/weak coupling duality $b \to 1/b$.   Since there is no understanding nor evidence for such a duality based on the quantum action of the model,  it should be questioned whether the properties of the model for $b>1$ are simply obtained by analytic continuation of the weak coupling regime $0<b<1$. In this article we assert that the answer is no,  and we develop a concrete and specific proposal for the properties when $b>1$. Namely, we propose that in this region one needs to introduce a background charge $\Q = b + 1/b -2$ which differs from the Liouville background charge by the shift of $-2$. We propose that in this regime the model has non-trivial massless renormalization group flows between two different conformal field theories. This is in contrast to the weak coupling regime which is a theory of a single massive particle. Evidence for our proposal comes from higher order beta functions. We show how our proposal correctly reproduces the freezing transitions in the multi-fractal exponents of a Dirac fermion in $2+1$ dimensions in a random magnetic field,  which provides a strong check since such transitions have several detailed features.  We also point out a connection  between a semi-classical version of this transition and the so-called Manning condensation phenomena in polyelectrolyte physics.
\end{abstract}

\maketitle

\vskip -0.8 truecm

\tableofcontents

\section{Introduction}

The sinh-Gordon model is the simplest relativistic model in $1+1$ dimensions that is integrable.    
It can be defined by the action 
\beq
\label{Action}
\CS = \int d^2 x \( \inv{8 \pi}  \d_\mu \phi \d_\mu \phi + 2 \mu \, \cosh(\sqrt{2} b \phi) \),
\eeq
with $b$ is a real parameter and $\phi$ a real scalar field. 
The current understanding is that the  spectrum consists of a single massive particle,  with an S-matrix that is factorizable in terms of the
two particle S-matrix \cite{sinhG1},  eqn. \eqref{Smatrix} below.   Based on this,  a great deal is known about the model.   
As a partial list of references,  let us mention the following.    
The form factors have been computed \cite{MussardoSinh1,MussardoSinh2},   which can be used to calculate correlation functions.     Even finite temperature 1-point correlations are computable \cite{LeClairMussardo}.   
The thermodynamic Bethe ansatz (TBA) has also been investigated \cite{ZamoTBAsinhG};  we will say more about this later.  

In spite of this vast amount of known results concerning the sinh-Gordon model,  one important aspect has essentially remained unanswered.     Remarkably,  the S-matrix satisfies the strong/weak coupling duality in that it is invariant under $b \to 1/b$.    
The most common viewpoint is that one first defines the theory for $0<b<1$  where one can trust perturbation theory around $b=0$, and then one {\it defines} the theory 
for $1<b<\infty$ using the duality.    Since the form factors and TBA are invariant under $b \to 1/b$,  from this perspective  the theory for $b>1$ is the same as the dual theory at $0<b<1$.     
However an important question arises.    Since there is no indication of a $b \to 1/b$ duality based on the action 
\eqref{Action} itself,   the analytic continuation $b \to 1/b$ may actually not be valid.   
Relatively recently this issue was studied in much detail using a truncated Hilbert space approach \cite{KonikMussardo} and indeed indications were found that this approach breaks down as $b$ approaches the self-dual point $b=1$.    
It was suggested there that for $b>1$ the theory may actually be massless,  however definite properties of such a theory remained unspecified and are still unknown.  A similar breakdown of analyticity is known to occur in freezing phenomena, such as in disordered systems \cite{Derrida}, and was recently shown to be present in Coulomb gas systems \cite{BouchaudFyodorov} which bear analogies with the sinh-Gordon model. The freezing transition in the sinh-Gordon, in connection with disordered fermions, was also considered in \cite{Doussal,Doussal2}. We will say more about this connection  later.

The purpose of the present article is to make a definite proposal for the behavior of the sinh-Gordon model for $b>1$ that is not a simple analytic continuation $b \to 1/b$ of the $0<b<1$ regime.    
Our specific proposal is easily described.   For $0<b<1$ the theory can be defined as a perturbation of a free massless boson in the ultraviolet (UV), and the standard properties based on a single massive particle with the known S-matrix all apply.  
However at $b=1$ and above,   a background charge $\Q$ is spontaneously generated.    This background charge is not the same as one would obtain if one views the sinh-Gordon theory as a perturbation of the Liouville conformal field theory (CFT), 
but is rather given by 
\beq
\label{Qinf}
\Q = b + 1/b -2
\eeq
which is a shift of the Liouville value by $-2$.     
Furthermore,  the $b>1$ regime is a massless phase,   but is not conformally invariant.    Rather the theory can be described by a massless renormalization group (RG) flow  between two conformal field theories.   

In Section \ref{shGbc} we propose our main result \eqref{Qinf} based on some rather simple criteria.    
However these  simple arguments by themselves are not enough to understand the true nature of the theory beyond the self-dual point.    Ultimately the properties of this theory should be tied to properties of the RG,  since the non-zero background charge 
$\Q$ affects anomalous dimensions,  etc.     To this aim we study the RG for the sinh-Gordon model based on the 
beta functions proposed in \cite{Moriconi} and understood in more detail in \cite{BL}.     These beta functions are well suited to our purposes since they are ultimately  based on the action  \eqref{Action} of the sinh-Gordon theory itself. 
More importantly,   the physics of interest here concerns massless flows between conformal field theories,   and 
it was shown in \cite{BL} that these proposed beta functions predict RG flows that agree precisely with exact results for massless flows in the so-called  ``imaginary" sine Gordon model \cite{FSZ1,FSZ2}.  We will review this below.  
These beta functions also correctly predicted cyclic RG flows.   
  This gives us some confidence in at least attempting  to use these beta functions to explore the physics we are trying  to understand.  
 As we will show  in Section
\ref{AllOrdersBeta} there is a clear difference between the RG flows for $b<1$ verses $b>1$ since these beta functions do not have the symmetry $b \to 1/b$.        
Furthermore,   we can argue based on these beta functions that $b>1$ is a massless phase and provide support for 
our proposed $\Q$.

The beta functions in \cite{Moriconi} are based on anisotropic current-current perturbations of a Wess-Zumino-Witten model at 
a level $k$,  and the map to sine and sinh Gordon theory was made in \cite{BL},  which only involves level $k=1$.  
It needs to  be mentioned that the beta functions in \cite{Moriconi} are still conjectural.    It was argued in \cite{Wiese} that there are corrections at 4-loops.    One of us has also pointed out that there could be $1/k$ corrections to these beta functions for a different class of  ``flavor" anisotropic models \cite{Chiral}.     
Similar kinds of  ``all-orders"  beta functions were considered in \cite{Greeks1,Greeks2, Greeks3,Tseytlin} using rather different gravitational methods.   There it was  also argued that there are higher $1/k$ corrections.    
 On the other hand,  as already stated above,   the renormalization scheme used in \cite{Moriconi} to obtain an all-orders beta function has already been shown  to provide  exact results for the kind of physics being explored here.      
In light of these statements,  in this paper we will simply assume the beta functions in \cite{Moriconi,BL} to be correct 
 enough  to capture the physics we are trying to understand and leave aside the issue of possible corrections and whether they affect our conclusions.     Our analysis of these beta functions at this stage should be viewed as supportive,  but not indisputable,  evidence for our main proposal described above.   
In any case,  irrespective of the present work, it is not  at all understood how these  proposed $1/k$ corrections can be reconciled with 
the correct exact predictions on massless flows in \cite{BL,FSZ1,FSZ2}.

In Section \ref{disorder} we apply our proposal to the freezing transition of a Dirac fermion in $2+1$ dimensions  in a random $U(1)$ gauge field, namely a magnetic field.
We first map the problem to the sinh-Gordon theory.   Then using our proposed $\Q$, we compute in detail the multi-fractal exponents and their transitions,  which can all be traced to the transition at $b=1$ of the sinh-Gordon model.    
Our results agree with known results based on Derrida's random energy model or other random fermion models \cite{Derrida,Castillo,Doussal,Doussal2}.   
This provides rather strong evidence for our proposals concerning the  transition in the sinh-Gordon model at the self-dual point.    

By studying a simple  semi-classical evaluation of one-point functions in the sinh-Gordon model,  one can understand how the premises of a transition can be found in such an approximation,  and how it is related to the well-known counter-ion Manning's condensation \cite{Manning} in polyelectrolyte solutions. This semi-classical computation actually points towards a freezing transition in the spectrum of possible exponential operators, as a function of their weights for fixed value of the sinh-Gordon parameter $b$.
We relegated these results to an  Appendix since the calculation is semi-classical and detailed properties are beyond its scope.  
However we found it instructive to include this analysis since it provides a simple intuitive picture for the transition.

\section{Sinh-Gordon conventions}

\label{sinhGconventions}

Since there are several conventions in the literature,   and factors of $1/4 \pi$  and $\sqrt{2}$ are important here,  
we clearly define our conventions.  
In the standard understanding,   almost certainly valid in the weak coupling regime $0<b<1$, 
the spectrum consists of a single particle of mass $\mphys$ with two particle S-matrix
\beq
\label{Smatrix}
S(\theta) = \frac{ \sinh \theta - i \sin \pi \gamma }{\sinh \theta + i \sin \pi  \gamma}, ~~~~~~~\gamma \equiv b^2/(1+ b^2).
\eeq
Here $\theta$ is the difference of the usual rapidity parameterization of energy/momentum:  
$(E,p) = \mphys \, (\cosh \theta, \sinh \theta )$. This S-matrix is invariant under the duality $b\to 1/b$ which  corresponds to  $\gamma\to 1-\gamma$.

The free  gaussian field when $\mu =0$ can be decomposed as $\phi (z, \zbar) = \varphi (z) + \bar{\varphi} (\zbar)$, where $z= x+i y$ and $\zbar = x- iy$.  
With the $1/8\pi$ in the action,  the above fields have the canonical two point functions, $\langle \varphi (z) \varphi (w) \rangle = - \log (z-w)$, and similarly for $\bar{\varphi}$,  and its Virasoro central charge is $c=1$.       It is most natural to view the $\cosh$ potential as a perturbation of the free  gaussian field.    
Let $\dim{*}$ denote the total scaling dimension of $*$ in mass units (for fields this is the sum of left and right  conformal dimensions
$\Delta + \bar{\Delta}$).   
One has
\beq
\label{dimcosh}
\dim {\cosh( \sqrt{2} b \phi )} = -2 b^2,
\eeq
which is always relevant for real $b$.    
Thus $\mu = [{\rm mass}]^{2+2b^2}$ for some mass parameter.      One can take the latter as the physical mass of the single sinh-Gordon particle, such that 
\beq
\label{mphysmu}
\mphys = F(b) \, \, \mu^{1/(2 + 2 b^2)} .
\eeq
The non-trivial function $F(b)$ was computed by Al. Zamolodchikov by comparing conformal perturbation theory with the thermodynamic Bethe ansatz  (TBA) since the latter is expressed in terms of $\mphys$ \cite{ZamoMassScale}.  We will not need the explicit form of $F(b)$ which is somewhat complicated  but only its value in the limit $b \to 1$.

We can now clearly address the issue we are proposing to resolve in this paper that was referred to in the Introduction. 
The S-matrix \eqref{Smatrix} is invariant under the strong/weak duality $b \to 1/b$.     The self-dual point is $b=1$.    This,  combined with the  $Z_2$ symmetry $b \to -b$ naively suggests one need only solve the theory the region $0\leq b \leq 1$,  and then analytically continue the result to all $b$ on the real line.    First of all,  there is no guarantee this analytic continuation is valid since it is not a symmetry of the lagrangian whatsoever.   Moreover there is concrete  evidence that some phenomenon  is going on at $b=1$ which as yet is not understood.     One indication is that from the exact form of $F(b)$,  one finds
\beq
\label{mto0}
\lim_{b \to 1}  \, \mphys = \lim_{b \to 1} \,\,\frac{4  \sqrt{\pi}}{\Gamma \(\tfrac{1}{4}\)^2 } \( \frac{\pi \mu}{\Gamma (1-b^2) } \)^{1/4}  =0, 
\eeq
which formally implies $\mphys =0$ at $b=1$.   For $b>1$ it would appear $\mphys$ is complex.   If the physical mass is zero at $b=1$,  then the S-matrix \eqref{Smatrix} does not make much sense since rapidity is not defined if $\mphys =0$.     More recently the sinh-Gordon model was studied starting from the action  in a truncated Hilbert space approach and clear deviations from the TBA predictions were observed as $b\to 1$ \cite{KonikMussardo}.    On the other hand, for $b$ not too large but still in the region $0\leq b \leq 1$,  results based on a truncated space of the free gaussian field work very well. Based on this one could conclude that the theory defined by the lagrangian   for $b>1$ has different properties than one would expect from an analytic continuation of $0 \leq b < 1$. Another piece of evidence for this transition comes from  the analysis of random fermions done in \cite{Doussal,Doussal2}.  There, the existence of  a freezing transition in the sinh-Gordon was conjectured in connection with glassy behavior of random Dirac fermions. Their analysis was based on using one-loop RG equations, in the form of the so-called KPP equations, and their traveling wave solutions.
 Up to now these works, though very interesting,    have not  provided a  concrete indication  of the properties of the theory for $b>1$.

\section{Sinh-Gordon theory with a background charge}
\label{shGbc}

\subsection{Generalities}

In this section we consider the sinh-Gordon model with a background charge $\Q$ at infinity.   Formally one can deform the action as follows:
\beq
\label{SQ}
\CS = \CS_{\rm shG} \, + \, \Q \, \phi_\infty .
\eeq
Alternatively one can couple the field $\phi$ to the curvature $R$,   adding a term proportional to $\Q\, R\, \phi$ to the lagrangian.   
Either way,  in the unperturbed $\mu =0$ conformal field theory,   the effect is to modify the conformal stress tensor 
\beq
\label{TQ}
T (z) = -\tfrac{1}{2} \d_z \varphi \d_z \varphi + \tfrac{\Q}{\sqrt{2}}  \,\,  \d_z^2 \varphi ,
\eeq
and similarly for $\bar{T} (\zbar )$.
The Virasoro central charge is now
\beq
\label{Vir}
c = 1 + 6 \,\Q^2 .
\eeq
The main effect of non-zero $\Q$ is to change the scaling dimensions of operators in the free boson CFT~:
\beq
\label{DimQ}
\dim{e^{\sqrt{2} a \phi}} = 2 a (\Q-a).
\eeq
The two exponentials in the $\cosh$ now have different scaling dimensions,  thus one should write
\beq
\label{coshExp}
2\mu \, \cosh (\sqrt{2} b \phi ) ~~~ \to ~~ \mu_+ e^{\sqrt{2} b \phi} + \mu_- e^{-\sqrt{2} b \phi}.
\eeq
Although the dimensions of $e^{\pm \sqrt{2} b \phi}$ differ, their sum adds up to $-4b^2$ for any $\Q$.

For a weight ``$a$"  in $e^{\s2 a \phi}$, let us define its dual $\adual$~:
\beq
\label{adual}
\adual \equiv  \Q-a ,
\eeq
and similarly for the sinh-Gordon coupling $\bdual \equiv \Q -b$.     
Note that for zero $\Q$,  $\adual = -a$,  including of course $\bdual = -b$.   Thus with no background charge, the duality 
$b \to \bdual$ simply corresponds to the $Z_2$ symmetry of the action $b \to -b$.    

For the CFT with non-zero $\Q$, one has the duality that the dimension of $e^{\s2 a \phi}$ is invariant under $a \to \adual$.     Coulomb gas techniques indicate the equivalence 
\beq
\label{Coulomb}
e^{\s2 a \phi} ~ \simeq ~e^{\s2 (\Q-a) \phi},
\eeq
in the CFT  correlation functions. 
In the TBA this equivalence can be expressed in terms of so-called reflection amplitudes $R(a)$
\beq
\label{Reflection}
e^{\s2 a \phi} ~ = R(a) \, e^{\s2 (\Q-a) \phi}
\eeq
which are known for the Liouville case \cite{LiouvilleShG2}. This reflection symmetry is known to be valid in the Liouville theory but only conjectural in the sinh-Gordon theory (see the Appendix for a discussion of this point).

For any background charge $\Q$,  the effective central charge $c_{\rm eff}$ of the TBA is the same if the particle is considered massive.      The TBA equations based on the S-matrix \eqref{Smatrix} do not depend explicitly on $\Q$.   However the 
effective UV central charge is $c_{\rm eff} = c_{\rm vir} - 12 d_0$ where $c_{\rm vir}$ is the Virasoro central charge 
and $d_0$ is the ground state energy.   Now $d(a) = 2 a (\Q-a)$ which has a maximum at $a=\Q/2$ which corresponds to $d_0=\Q^2/2$.   Thus $c_{\rm eff} = 1 + 6 \Q^2 -12 (\Q^2 /2) = 1$, independently of $\Q$.  

In principle,  the sinh-Gordon model can be considered with any $\Q$.     In this paper we will  only consider two choices,   the Liouville case and the choice described in the subsequent subsection. Let us consider the first.

\subsection{Liouville case} 

This is the most natural choice besides the perturbation of the $\Q=0$ free massless boson.    
Many works indicate that the sinh-Gordon model may be viewed as a perturbation of the Liouville theory, in particular  \cite{LiouvilleShG1,LiouvilleShG2,LiouvilleShG3}.   
In this choice, 
\beq
\label{QLiouville}
\Q = b + 1/b ,  ~~~~\Longrightarrow ~~ \dim{e^{\s2 b \phi}} =2.
\eeq
Namely the positive exponential is an exactly marginal operator,   $\dim{\mu_+} = 2$,  and the additional $e^{-\s2 b \phi}$ is viewed as a relevant perturbation of the Liouville CFT.    Although this may seem like an unnecessary complication,  surprisingly  it has been shown that the conformal perturbation theory of this model with non-zero $\Q$ can reproduce the perturbation theory with $\Q=0$ when $0<b<1$ \cite{KonikMussardo}.       

For $\Q = b + 1/b$, the duality \eqref{adual} is $b\to \bdual = 1/b$.
This indicates that this Liouville formulation of the sinh-Gordon model is unable to address the problem posed in this paper since this dual is the usual one that maps 
the region $0\leq b \leq 1$ to $1 \leq b \leq \infty$.     Thus it has nothing to say about any novel behavior for $b>1$.   

\subsection{Freezing  transition at the self-dual point}

Our  aim is to find a different choice of background charge $\Q$ that can define the sinh-Gordon model for $b>1$,   which is expected to  have different  properties.     
As discussed above,  the model with $|b|\leq 1$ appears to be well-defined as a perturbation of the free massless boson with zero $\Q$, or as a perturbed Liouville theory.  
Taking the simpler option,  we assume there is no background charge in this region,  i.e. 
\beq
\label{Qzero}
\Q = 0, ~~~~{\rm {for} ~~  0<b< 1} .
\eeq
At $b=1$ we introduce a non-zero $\Q$.      It would be very interesting to understand what precise mechanism spontaneously generates this non-zero $\Q$,   however we leave aside that question in this work.     

\bigskip
\noindent
The conditions we impose on $\Q$ for $|b|>1$ are quite natural and are the following~:

\begin{itemize}
\item
Based on the symmetry of the S-matrix,  we require $\Q$ to be self-dual, as for the Liouville case.    This implies we can expand $\Q$ as a series in $b + 1/b$~: $\Q = \sum_{n=0}^\infty \alpha_n \(b+1/b\)^n$.
Since there are not enough constraints to fix all $\alpha_n$,   we assume only $\alpha_0$ and $\alpha_1$ are non-zero,   as in the Liouville theory.    

\item
For continuity with $|b|<1$,  we require the background charge $\Q=0$ at $b=1$~: $\Q = \alpha_0 + \alpha_1 (b+1/b) = 0$ at $b=1$, so that $\alpha_0 = - 2 \alpha_1$.
This in turn implies that at the self-dual point $b=1$,  in the UV one has  $c=1$,  as for $0<b<1$.      Thus in the UV,  the central charge $c$ is continuous and only changes at the self-dual point $b=1$.     

\item
To fix $\alpha_1$ we need a condition at $b= \infty$.    We require that under the duality $b \to \bdual = \Q -b$,  the dual coupling constant $\bdual$ remains in the non-zero $\Q$ region 
$|\bdual | \geq 1 $.   For a fixed $b$, this amounts to  $ \bdual   \leq  -1$, which implies $\alpha_1 \leq b/(b-1)$. 
Requiring the above for all $b$, in particular $b=\infty$,  leads to the minimal choice  $\alpha_1 = 1$.    
Although at this stage this appears somewhat ad hoc,  as we will see it leads to correct predictions for the random energy model.  
\end{itemize}

In summary,  we thus propose 
\beq
\label{Qfinal}
\Q = b + 1/b -2 ,~~~~{\rm {for} ~~  b > 1} ,
\eeq
and zero otherwise.  
This is just a shift of the Liouville background charge by the integer $-2$.   
Note that as $b \to \infty$,  $\Q$ is the same as for the Liouville case.  
Notice  also the dichotomy:  For $\Q=0$ and the Liouville choice $\Q=b+1/b$,   $\bdual = -b$ and $1/b$ respectively,  whereas for the above choice $\bdual = -2 + 1/b$, which equals $-b$ for $b=1$.  

With this choice of $\Q$ one has 
\beq
\label{dims}
\dim{e^{\s2 b \phi } }= 2 - 4 b, ~~~~~~
\dim{e^{-\s2 b \phi} }= 4b -2 - 4 b^2 .
\eeq
Thus the dimensions of the parameters $\mu_\pm$ are 
\beq
\label{dims2}
\dim{\mu_+} = 4b, ~~~~~~
\dim{\mu_-} = 4(b^2-b+1). 
\eeq
For $b \geq 1$, both operators $e^{\pm \s2 b \phi}$ are then relevant, even though they have different dimension.      The ultra-violet limit is controlled by the highest dimension operator,  namely the least relevant.
This  is the operator $e^{+\s2 b \phi}$.    We thus propose that in this frozen phase $b>1$ one effectively has $\dim{\cosh ( \s2 b \phi )} \sim 2 - 4 b$.    The term ``frozen" is borrowed from the theory of disordered systems;  see below.  

In summary,  we have proposed that 
\beq
\label{dimcases}
\dim{e^ {\s2 b \phi}} = 
\begin{cases} 
-2b^2, ~~~~~~&{\rm for} ~~0< b < 1, \\
2 - 4 b, ~~~~&{\rm for} ~~ b>1 .
\end{cases}
\eeq
Furthermore,  as we explained,  we identify the above dimensions \eqref{dimcases}  as the effective scaling dimension of $\cosh ( \s2 b \phi )$.   
This transition is induced by the generation of a background charge for $b>1$.   We conjecture that the sinh-Gordon model, which is well defined for $0<b<1$ with $\Q=0$, is actually ill-defined for $b>1$ without background charge but well-defined with the background charge $\Q = b + 1/b -2$ as in eq.\eqref{Qfinal}.

\section{Massless Renormalization group flows in the sine- and  sinh-Gordon models}

In the present context by ``massless flows" we mean the following.   Suppose an RG flow originates as a perturbation of an UV fixed point CFT by a relevant operator of dimension 
$\Gamma_{UV} <2 $ 
and flows to a different non-trivial fixed point in the infrared  (IR),  necessarily arriving there via an irrelevant operator of dimension $\Gamma_{IR}>2 $.    Generally in the flow to the IR,  massive particles decouple, thus if the IR theory is non-trivial some massless particles must survive the flow.    In the deep IR, the theory is approximated by the interactions of these massless degrees of freedom.    

\subsection{Higher order  beta functions}
\label{AllOrdersBeta}

The sinh-Gordon model can be viewed as a current-current perturbation of an SU(2) WZW  model at level $1$ with action,
\beq
\label{WZWpert}
\CS = S_{{\rm wzw}} + \inv{2 \pi} \int d^2 x \( g_1 \big[J^+ \bar{J}^- + J^- \bar{J}^+ \big] + g_2 J_3 \bar{J}_3 \),
\eeq
with $J^\pm = \tfrac{1}{\sqrt{2}} e^{\pm i \sqrt{2} \varphi }$ and $J_3 = \tfrac{i}{\sqrt{2}} \d_z \varphi $, 
where $\varphi(z)$ is the $z$-dependent part of a free massless scalar field $\phi(z, \zbar) = \varphi (z)+ \bar{\varphi} (\zbar) $.
The advantage of doing this is that current algebra Ward identities allow an easier approach to calculating higher order corrections to the beta functions since both couplings $g_{1,2}$ are marginal.    
The bosonized form of the action is now 
\beq
\label{bosonize2}
\CS = \inv{4 \pi} \int  d^2 x \( \half (\d \phi)^2 + g_1 \, \cosh (\sqrt{2}  b  \phi ) \). 
\eeq
The coupling $b$ is a function  of $g_2$ presented below,  and can be  real  or  imaginary corresponding to  either  sinh  or  sine  Gordon phases \cite{BL}.  For reasons that will become clear,  let us postpone this identification for now since such an identification depends on $\Q$, and first describe the general properties of the flows based solely  on the beta functions for $g_1, g_2$.   

In our original treatment \cite{BL},  $g_1$ was taken to be real.    For several reasons in this section we present our conclusions for $g_1$ imaginary.     One reason is that for our purposes we are interested in massless RG flows,  and this case provides a point of comparison with the known exact results \cite{FSZ1,FSZ2}.    The second is that for the map to sinh-Gordon for disordered systems, equation 
\eqref{Action3} below,  $g_1$ is indeed imaginary.    Thirdly,  under the continuation $g_1 \to i g_1$,  the poles in the beta functions in \cite{BL} no longer exist and one does not have to deal with continuing the flow through these poles.  
Fortunately as we will comment on at the end of this section,  for the sinh-Gordon flows we are interested in the distinction between real and imaginary $g_1$ does not matter as far as the endpoints of the flows are concerned,  even though the details of the  RG trajectories do differ.   

For  the reasons just described above,  we  extend the results in \cite{BL}  to $g_1 \to i g_1$, corresponding to imaginary $\mu$ in \eqref{Action}.   
 Taking $g_1 \to i g_1$,  the  beta functions
in \cite{Moriconi} become the reasonably simple functions
\begin{subequations}
\begin{align}
\label{betas}
\frac{dg_1}{d \log a } &\equiv  \beta_{g_1} = \frac{g_1 (g_2 + g_1^2/4)}{(1 + g_1^2 /16)(1+g_2 /4)}~, \\ 
\frac{dg_2}{d \log a } &\equiv \beta_{g_2} = - \frac{g_1^2 (1-g_2/4)^2 }
{( 1 + g_1^2/16)^2} ~,
\end{align}
\end{subequations}
where $a$ is a cut-off scale.    With these conventions, the flow to the IR corresponds to $a\to \infty$.  
Flows with $g_1 <0$ are just a mirror image of those with $g_1>0$ since the beta functions are invariant under $g_1 \to - g_1$,  thus we will only discuss the case $g_1>0$.    

The above beta functions have a remarkable strong/weak coupling duality.   For both $\beta_{g_1, g_2}$ (recall that beta functions transform as vector fields)~:
\beq
\label{betaDuality}
\beta(16/g_1, 16/g_2) = - \beta (g_1, g_2).
\eeq
Also,  again rather remarkably,   there exists an RG invariant which allows us to map out  basic features of the flows without explicitly solving the coupled differential equations based on the beta functions.  
Such an invariant $\II$ satisfies $\sum_g \, \beta_g \, \d_g \II = 0$.
One may check that the following $\II$ is an invariant \cite{BL}
\beq
\label{Q}
\II(g_1, g_2)  = \frac{g_1^2 + g_2^2}{(g_2 -4)^2 (g_1^2+ 16)} . 
\eeq
This invariant satisfies the strong-weak coupling duality of the beta functions \eqref{betaDuality},
\beq
\label{duality}
\II(16/g_1, 16/g_2) = \II(g_1,g_2). 
\eeq

The line $g_1=0$ is a line of fixed points where both beta functions are zero.  
By computing the slope of the beta function near $g_1 =0$,  one can determine the dimension $\Gamma_0$ of the perturbation  $\cosh (\sqrt{2} b \phi)$ there with the general formula
\beq
\label{slope}
\beta (g) = (2-\Gamma(g_c)) (g-g_c) + \cdots,
\eeq
near a critical point $g_c$. Since $\beta_{g_1} =  \[ 4 g_2/(4 + g_2 )\] g_1$ near $g_1 = 0$, equating the slope with $2 - \Gamma_0 (g_2) $ yields
\beq
\label{Gamma0}
\Gamma_0 (g_2)  \equiv \Gamma(g_1=0) =  2 \( \frac{4-g_2}{4+g_2} \) .
\eeq
Thus, the $g_2$ axis  (at $g_1$ =0) is divided into several  regions,  where the perturbations are classified as relevant $(\Gamma_0 < 2)$ or irrelevant  $(\Gamma_0 > 2)$.   We thus identify three distinction regions at $g_1=0$~:

\medskip

\noindent
-- {Relevant~: ~~~~}  $-\infty <g_2 < -4$;\\
-- {Irrelevant~: ~}~~  $-4 < g_2 <0 $;\\
-- {Relevant~: ~~~~}  $\ 0<g_2 < \infty$.
\medskip

\noindent
For the relevant regions the flows {\it originate} at $g_1=0$ and flow toward increasing $g_1$.    
For the irrelevant region,  the flows {\it terminate} at $g_1 =0$ arriving from positive $g_1$.

As we will see,  based on the invariant $\II$,  many of the flows that originate at $g_1=0$ end up at $g_1 = \infty$. 
Not all however,  depending on whether $g_1$ is real or imaginary,  see below.   But in the case of interest, namely the sinh-Gordon model,  flows indeed start at $g_1 =0$ in the UV and flow to $g_1 = \infty$ in the IR.   
Whereas the dimensions of the perturbations around $g_1 =0$ are unambiguous as a function of $g_2$ 
 since $g_1$ is an obvious line of fixed points (see $\Gamma_0 (g_2)$ in \eqref{Gamma0}),  
 the scaling dimension of the perturbation at $g_1 = \infty$ is less obvious.     We propose the following identification.   
Based on the duality of the beta functions \eqref{betaDuality},  flows at $g_1 =0$ can be mapped into flows at $g_1=\infty$,   where formally if the beta functions are zero at $g_1 =0$ they are also zero at $g_1=\infty$ if one uses $(16/g_1, 16/g_2)$ as coordinates.     However the minus sign in \eqref{betaDuality} implies the UV and IR are exchanged,  since they are related by $a \to 1/a$.  
We propose that along the flow the dimension of the perturbation is given by the same functional form as $\Gamma_0 (g_2)$ for $g_2$ as a function of the RG scale.  To be more precise, we are assuming that the dimension of the perturbation $\Gamma (t)$,  where $t=\log a$ is the RG time, 
satisfies $\Gamma(t) = \Gamma_0 (g_2 (t))$.
Thus at $t=0$, $\Gamma = \Gamma_0 (g_2 (0)) = \Gamma_0 (\gUV) = \Gamma_{\rm UV}$. On the other hand  at $t= \infty$,  
\beq
\label{GammaIR}
\Gamma = \Gamma_0 (g_2 (\infty))   \equiv  \Gamma_{\rm IR} .
\eeq
 The flow of $g_2(t)$ thus implies a relation between $\Gamma_{\rm UV}$ and $\Gamma_{\rm IR}$.  
  
For instance if a flow originates at $g_1 = 0$ from a relevant perturbation with $\Gamma_{\rm UV} = \Gamma_0 (\gUV)$,  
and $\gUV$ flows to $\gIR$, 
then in the IR at $g_1 = \infty$ we identify the dimension as $\Gamma_{\rm IR}$  where 
\beq
\label{GIR}
\Gamma_{\rm IR} = \Gamma_0 (\gIR ) = 2 \( \frac{4-\gIR}{4+\gIR} \), ~~~ \mathrm{at}\ g_1 = \infty.
\eeq

Let us make several remarks supporting our  rather natural proposal \eqref{GammaIR} since it will be essential in the following~:  

\begin{itemize}
\item
The equation \eqref{GammaIR}  correctly predicts the exact relation between $\Gamma_{\rm UV}$ and $\Gamma_{\rm IR}$ for massless flows that both begin and end at $g_1 =0$.  These are  the flows in \cite{FSZ1,FSZ2}, see below.

\item
At $g_1 =\infty$,  the coupling $g_2$ stops flowing,  i.e. remains constant,  as it does at $g_1 =0$,  which implies the dimensions 
$\Gamma_{\rm UV}$ and $\Gamma_{\rm IR}$ are constant there.     One can see this as follows.   
One can eliminate $g_1$ and write the beta function in terms of $g_2$ and $\II$ only:
\beq
\label{betag2Q}
\beta_{g_2} = \frac{ 16\(g_2^2 - 16\II(g_2-4)^2 \)\(1-\II(g_2-4)^2\)}
{(g_2+4)^2}
\eeq
Now one has 
\beq
\label{Qlines}
\II_0 \equiv \II( g_1=0, g_2) = \frac{g_2^2}{16(g_2-4)^2}, ~~~~~~\II_\infty \equiv \II( g_1 = \infty, g_2) = \inv{(g_2-4)^2}.
\eeq
One sees that for both $\II=\II_0$ and $\II=\II_\infty$,   the beta function $\beta_{g_2} = 0$, thus $g_2$ is a constant in RG time both at $g_1 =0$ and $g_1 = \infty$. Alternatively, $\beta_{g_2}$ vanishes either for $16\II(g_2-4)^2=g_2^2$ or for $\II(g_2-4)^2=1$. The former corresponds to $g_1=0$, the latter to $g_1=\infty$.

\item
Since under the duality $g_2 \to 16/g_2$ the IR and UV limits are exchanged due to \eqref{betaDuality},  
one should expect that  $\Gamma_0 (16/g_2) = - \Gamma_0 (g_2)$,  which is satisfied.  

\item
When $g_2 = 4$,  $g_2$ does not flow at all since $\beta_{g_2} =0$ for all $g_1$. Thus it must be that 
$\Gamma_{\rm UV}  = \Gamma_{\rm IR}  = \Gamma_0  (4) =0$.    As we will see below,  this corresponds to the $b=0$ point of the sinh-Gordon theory which is just a free {\it massive}  boson.   
\end{itemize}

\noindent
Notice that at $g_1\to\infty$ in the IR, the theory might  either be the trivial massive theory, with all degrees of freedom frozen, or a non trivial theory, potentially a conformally invariant theory.

\subsection{Identification of sinh and sine Gordon phases}

It remains to identify where the above model with $g_1, g_2$ corresponds to the sinh-Gordon model.   This identification 
clearly depends on whether we assume the presence of a background charge or not, and this fact  will be important later.  
If we view the $\cosh$ potential as a perturbation of the free gaussian field with no background charge,  then $\Gamma_0 = -2 b^2$:
\beq
\label{bofg}
b^2 = \frac{(g_2-4)}{(g_2 + 4)} .
\eeq
Again -- at the price of repeating ourselves -- this identification relies on a specifically chosen relation between the scaling dimension and the parameter $b$ (which here assumes the absence of background charge).
If this dimension is positive we view the potential as being in a sine-Gordon regime $g_1 \cos (\sqrt{2} \beta \phi)$ with $\beta = i b$.   There are  now four distinct regions:
\medskip

\noindent
- {Relevant~: ~~}  $-\infty <g_2 < -4$, ~~~sinh-Gordon with $1< b^2 < \infty$;\\
- {Irrelevant~: }  $-4 < g_2 <0 $, ~~~~~~~\,sine-Gordon with $1 < \beta^2 < \infty$;\\
- {Relevant~: ~~} $ 0<g_2 < 4$, ~~~~~~~~~sine-Gordon with $0 < \beta^2 < 1$;\\
- {Relevant~: ~~}  $4<g_2 < \infty$, ~~~~~~~\,sinh-Gordon with $0 < b^2 < 1$.
\medskip

\noindent
The above regions are the same as those already identified in \cite{BL}.    Note already that the regions $b<1$ and $b>1$ are 
clearly distinguished.

\medskip

For $g_1 =0$ in the UV,  the  duality $g_2 \to 16/g_2$ corresponds to $b \to ib$,  i.e. maps from the sinh-Gordon to sine-Gordon regimes.    
On the other hand,  the usual hypothetical sinh-Gordon duality $b \to 1/b$ corresponds to $g_2 \to -g_2$.    However the latter is not a symmetry of the beta functions and indicates that the RG properties of $0< b < 1$ verses $b>1$ are indeed different.    
This is one of the main points of this paper which  we will subsequently explore in more detail.  

\subsection{Massless flows in the ``imaginary"  sine-Gordon model}

Following the terminology in \cite{FSZ1,FSZ2} we refer to the sinh-Gordon action \eqref{Action} with $\mu \propto g_1$ imaginary
as the ``imaginary" sine-Gordon model.
In this case there are flows that both begin and end at $g_1 =0$,   indicating a massless flow between two different CFT's,  both at $c=1$,  which differ in their radius of compactification $\beta$.   Here since the flows both start and end at $g_1=0$,  there is no ambiguity in determining anomalous dimensions in the UV nor the IR.   This situation was already explained in \cite{BL} based on the beta functions above,     however we review it here since it represents a prototype of the kinds of flows we will propose in the sinh-Gordon case. 

\begin{figure}[t]
\centering\includegraphics[width=.6\textwidth]{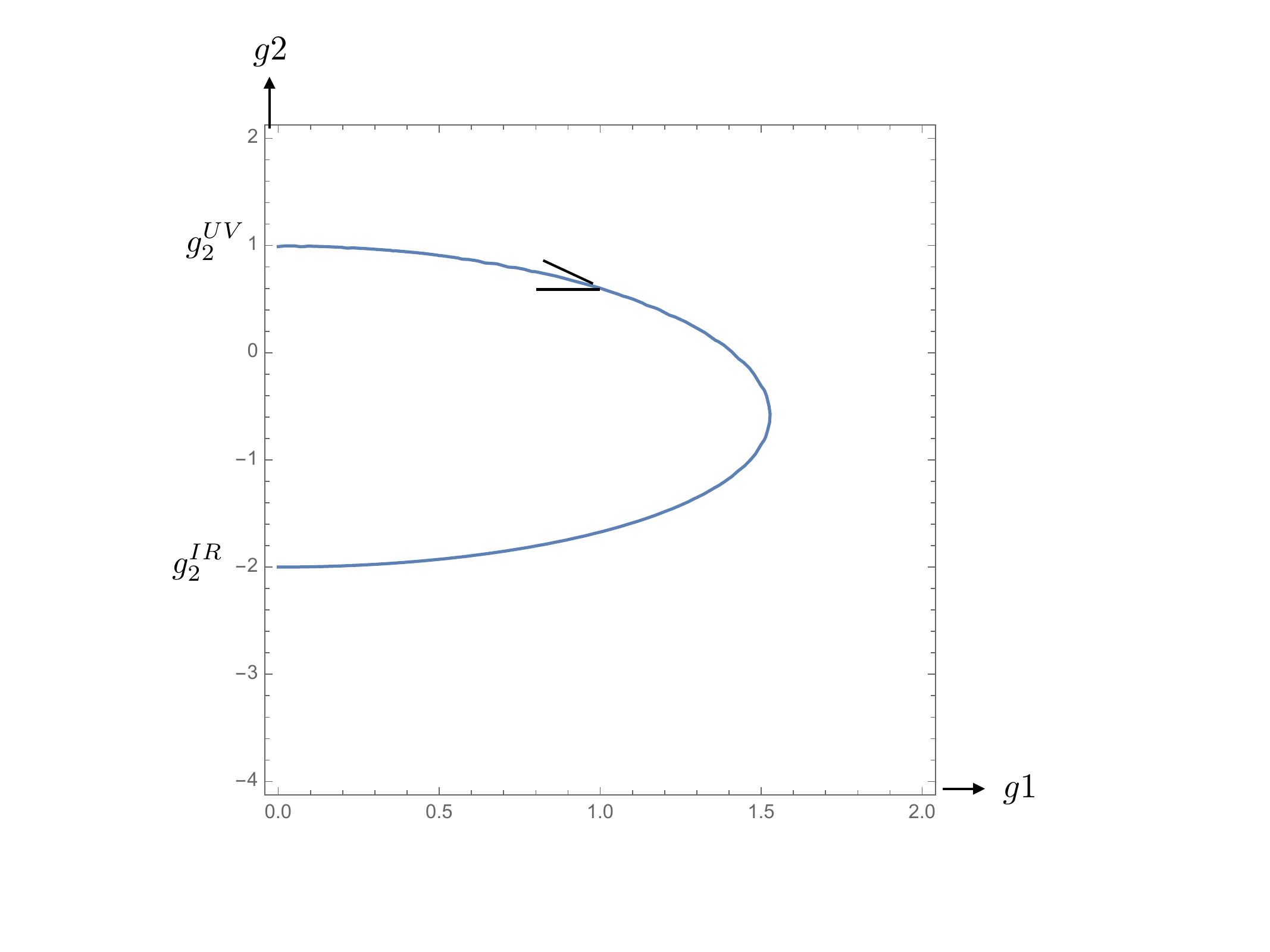}
\caption{Massless flow in the imaginary sine-Gordon model from $\gUV=1$ to $\gIR = -2$.     This corresponds to a flow from $\beta_{\rm UV}  = \sqrt{3/5} $ to $\beta_{\rm IR} = \sqrt{3}$.   What is shown is a contour plot of the RG invariant $\II$.  }
 \label{sineFig}
\end{figure}

In the sine-Gordon regime with small couplings $g_1$ and $g_2$,  $\II \propto (g_1^2 + g_2^2)$, thus the RG flows are approximately circles.    This implies that flows can both begin and end on the $g_2$ axis,   which is a massless flow as defined above.    Such flows are straightforward  to analyze to all orders.  For $g_1 = 0$,  
$\II=\II_0$ defined in \eqref{Qlines}.   Since $\II$ is preserved along the flow,  one must have 
\beq
\label{guvir}
\frac{\gUV}{\gUV -4} = - \frac{\gIR}{\gIR -4} ~~ \Longrightarrow ~~ \gIR = \frac{2 \gUV}{\gUV -2} . 
\eeq
In terms of $\beta$,
\beq
\label{betaUVIR}
\beta_{\rm IR}^2 = \frac{\beta_{UV}^2}{2\beta_{UV}^2 -1}, 
\eeq
which implies  the dimensions of the perturbation in the UV verses IR are related as follows
\beq
\label{UVIR}
\Gamma_{\rm IR} = \frac{\Gamma_{\rm UV}}{\Gamma_{\rm UV} -1 } .
\eeq
For  irrelevance in the IR, $\Gamma_{\rm IR} >2$,  requires $0< g_2 < 4/3$ or equivalently $1/2 < \beta^2 < 1$,  consistent with 
\cite{FSZ1, FSZ2}.      
A contour plot of such a flow is shown in Figure \ref{sineFig}.  
The existence of the flows and the relation \eqref{UVIR} have been conjectured long ago in \cite{FSZ1,FSZ2}. The fact that we recover them and the correct relation \eqref{UVIR} provides  further support for the effectiveness  of the beta functions \eqref{betas} in understanding  this kind of physics.  

\subsection{RG flows in the  sinh-Gordon model:  ~ $b<1$ verses $b>1$.} 

\begin{figure}[t]
\centering\includegraphics[width=.6\textwidth]{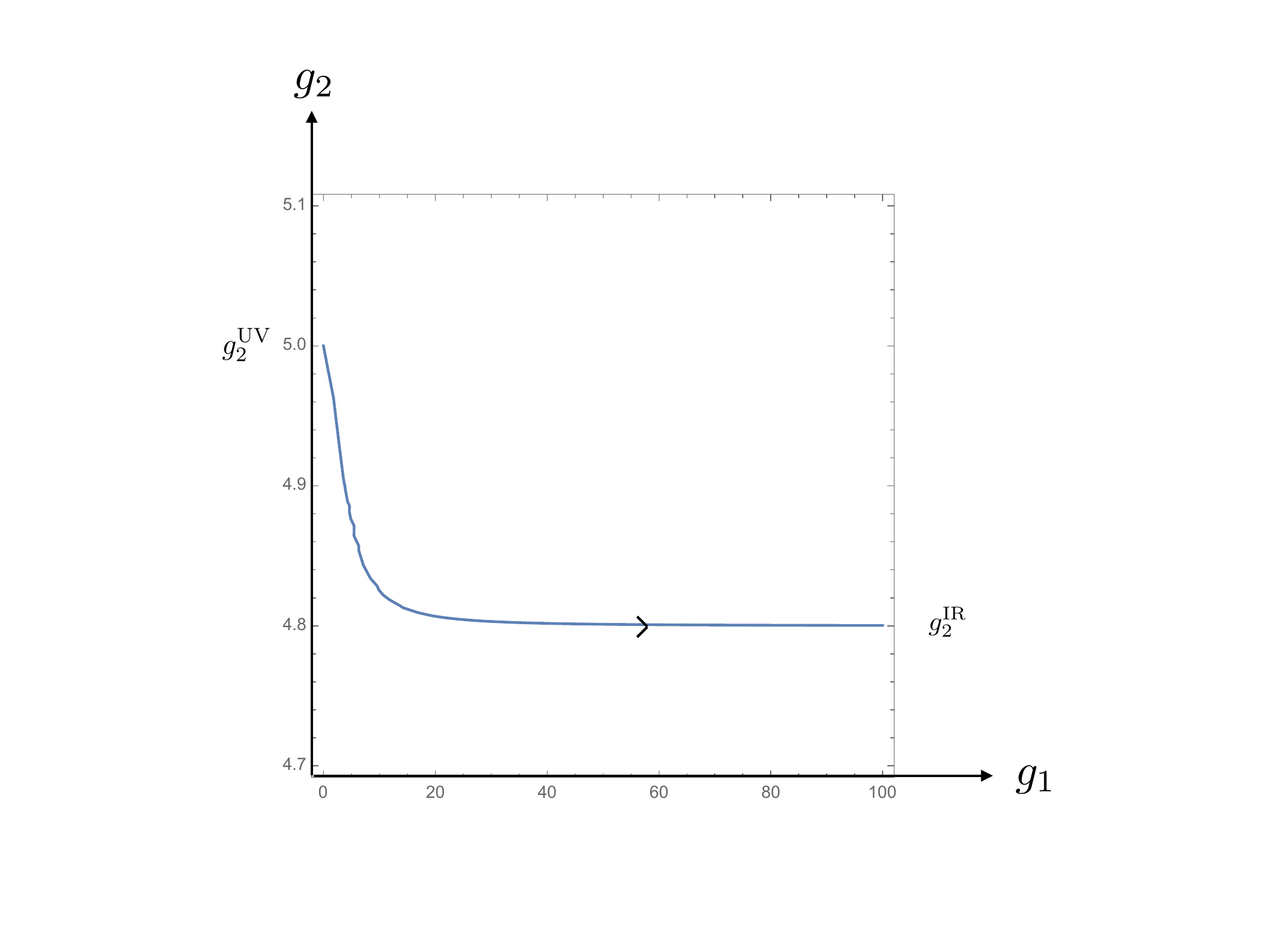}
\caption{A massive flow in the imaginary sine-Gordon model from  $\gUV = 5$ to $\gIR = 24/5 $.     This corresponds to a flow from $b_{\rm UV}  = 1/3$ to $b_{\rm IR} = 1/\sqrt{11}$.   What is shown is a contour plot of the RG invariant $\II$ where the $x,y$ axes corresponds to $g_1, g_2$.  }
 \label{sinhMassive}
\end{figure}

Here we consider flows in the different regimes $0<b<1$ and $1<b<\infty$ for the  sinh-Gordon with imaginary $\mu$.  
As explained at the end of this section,  the case of real $\mu$ is not very different.      Recall these regimes correspond to $g_2>4$ and $g_2 < -4$  at $g_1 =0$, respectively.  
  For such large coupling $g_2$,  constant $\II$ is not at all approximated by a circle as in the sine-Gordon case.   
 All flows originating at $g_1=0$ end up at $g_1 = \infty$.    As we now explain,  there are two cases which have rather different behavior, and correspond precisely to  $0<b<1$ verses $b>1$.  We need to relate $g_2$ in the UV and IR.  
 Since $\II$ is preserved along the flow, one must have $\II_0 (\gUV) = \II_\infty (\gIR)$.
A fortunate and promising result that has not been pre-programmed into the  above beta functions is that the two cases correspond precisely to weak verses strong coupling:   
 \medskip

 \noindent 
 $\bullet$ For {$0<b<1$}~: The flows originating at $(g_1, g_2) = (0, \gUV > 4) $ end up at $(\infty, \gIR))$  
 where 
 \beq
 \label{UVIRplus}
 \frac{\gUV}{4(\gUV -4)} = \inv{(\gIR - 4)} ~~~~\Longrightarrow ~~ \gIR = \frac{8 (\gUV -2)}{\gUV} .
 \eeq
 Expressing this in terms of the scaling dimensions~:
 \beq
 \label{Gammasplus}
 \Gamma_{\rm IR} = \frac{\Gamma_{\rm UV}}{1- \Gamma_{\rm UV}} ~.
 \eeq
Identifying the parameter $b$ using the relation $\Gamma=-2b^2$ yields $b_{\rm IR}^2= {b_{\rm UV}^2}/({1+2 b_{\rm UV}^2})$. Both $b_{\rm UV}$ and $b_{\rm IR}$ then remain in the weak coupling region $0<b<1$. This should be a massive flow since $\Gamma_{\rm IR}$  still signifies a relevant perturbation,   although with an imaginary coupling $\mu$.  One instance of it is shown in Figure \ref{sinhMassive}.
 \medskip
 
\begin{figure}[t]
\centering\includegraphics[width=.6\textwidth]{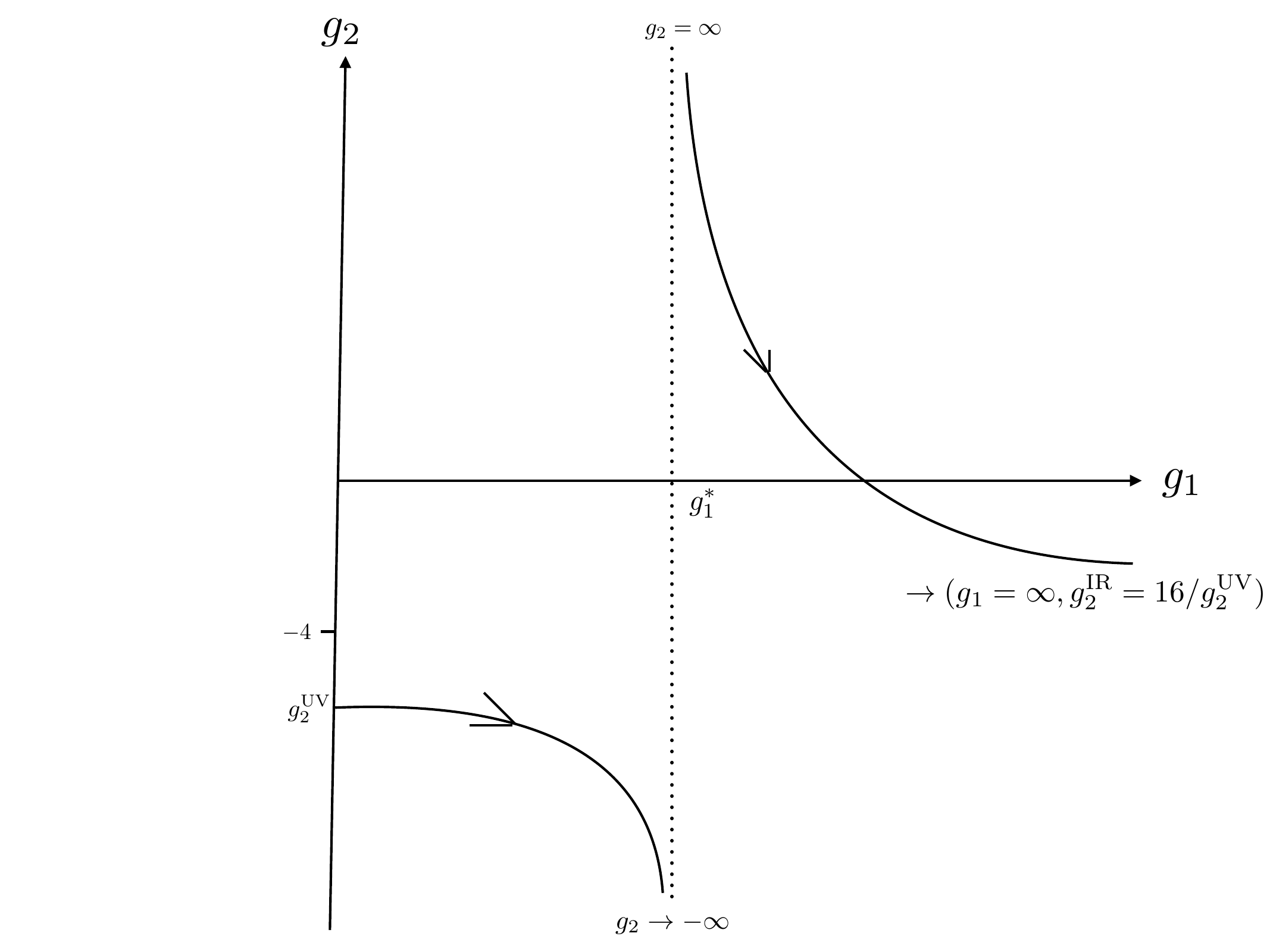}
\caption{Sketch of the massless flow in the imaginary sinh-Gordon model from $\gUV <-4 $ to $\gIR = 16/\gUV$.  
The value $g_1^*$ is given in the text: $g_1^* = {8\sqrt{4-2g_2}}/{|g_2|}$.   }
 \label{sinhFig}
\end{figure}

 \noindent 
 $\bullet$ For {$1<b<\infty$}~: The flows originating at $(g_1, g_2) = (0, \gUV <-4 ) $ end up at $(\infty, \gIR))$  
 where there is just a change of sign in \eqref{UVIRplus}:
 \beq
 \label{UVIRminus}
 \frac{\gUV}{4(\gUV -4)} = - \inv{(\gIR - 4)} ~~~~\Longrightarrow ~~ \gIR = \frac{16}{\gUV} .
 \eeq
 Expressing this in terms of the dimensions $\Gamma$:
 \beq
 \label{Gammasminus}
 \Gamma_{\rm IR}  = -\Gamma_{\rm UV} .
 \eeq
Now  in this case,  $\Gamma_{\rm IR} >2$,  i.e. irrelevant, and this is thus a massless flow.    
This flow in the sinh-Gordon region is rather  analogous to the massless flows in the sine-Gordon model described above,   since 
in the IR they both end up in the irrelevant regime where $-4< \gIR  < 0$. The details of the flow are however more intricate compared to the previous case.  Rather the flows start at $(g_1, g_2) = (0, \gUV)$   and  first flow to $g_2 = -\infty$.  This occurs at $g_1 = g_1^*$ such that $\II(0,g_2) = \II(g_1^*, \pm \infty)$, that is $g_1^* = {8\sqrt{4-2g_2}}/{|g_2|}$.
  Using the cylindrical topology proposed in \cite{BL} which identifies $g_2$ with $-g_2$ at $|g_2| = \infty$, the flow then continues  from $g_2 = \infty$ to $g_1 = \infty$ but with a different $g_2= \gIR$,  which is actually the dual of $g_2$.  This implies the flow 
\beq
\label{shGflow}
 (g_1, g_2) = (0,g_2) ~~~\UVtoIR  ~~~ (\infty ,16/g_2 ).
\eeq
 The self-dual point $b_{\rm UV} = 1$ flows to a marginally irrelevant perturbation in the IR, i.e. $\Gamma_{\rm IR} = 2^+$, which seems desirable if it is indeed a massless flow. Such flows are sketched in  Figure \ref{sinhFig} and were verified numerically.  
 
\medskip

We now explain why the above RG flows cannot be properly interpreted if we stick to the relation $\Gamma=-2b^2$, and argue that they acquire a natural interpretation if we introduce the background charge $\Q=b+1/b-2$. This is one of the main points of this paper.
\medskip

Let us present supporting arguments for the introduction of a background charge.
If we continue to identify the scaling dimension $\Gamma_0$ with $-2 b^2$  then the  above relation \eqref{Gammasminus}  implies the  peculiarity of $b_{\rm IR} = i b_{\rm UV}$, i.e.  becoming imaginary.  The flow to imaginary $b$ seems unsatisfactory  since it takes us out of the proper  sinh-Gordon regime manifold and into the sine-Gordon one.   
This would correspond, roughly speaking, from a flow from a non-compact model to a compact one.    We suggest that this problem arose since we identified the coupling $b$ with the dimension $\Gamma_0 = -2b^2$ which assumed there was no background charge.    Introduction  of the background charge $\Q$ in \eqref{Qinf} can resolve this issue.    

The perturbative calculations that led to the above beta functions \eqref{betas}  did not incorporate a background charge.   However the flows do predict dimensions of operators $\Gamma_{\rm UV, IR}$ regardless of the free gaussian identification $\Gamma = -2 b^2$.    Incorporating a background charge should just modify this identification,  while preserving the flows in $(g_1,g_2)$. We can indeed modify this identification,  but still must preserve the relation $\Gamma_{\rm IR} = - \Gamma_{\rm UV}$, since the latter is predicted by the beta functions regardless of the identification relating $\Gamma$ and $b$. 

We require that both the UV and IR are in the same regime of ``$b$".      
Let us identify $\Gamma_0$ with the dimension proposed in Section \ref{shGbc} 
\beq
\label{Gammabc}
\Gamma_0 = 2 - 4b 
\eeq
which was based on a background charge $\Q = b+1/b -2$ in the region $b>1$.  

The identification \eqref{Gammabc} modifies the relation between $g_2^\mathrm{UV}$ and $b_\mathrm{UV}$, as $4b_\mathrm{UV}=2-\Gamma_0(g_2^\mathrm{UV})$, which then reads $b_\mathrm{UV}=g_2^\mathrm{UV}/(4+g_2^\mathrm{UV})$. For $g_2^\mathrm{UV}<-4$, we still have $b_\mathrm{UV}$ in the strong coupling regime $b_\mathrm{UV}>1$.

Then $\Gamma_{\rm IR} = - \Gamma_{\rm UV}$ in \eqref{Gammasminus} implies the simple relation 
\beq
\label{bsIRUV}
b_{\rm IR} = 1 - b_{\rm UV}.   
\eeq
This has the desired property that the whole region  $b_{\rm UV} >1$ is mapped to $b_{\rm IR} <  0$ which excludes the usual sinh-Gordon region $0 < b < 1$.      
Importantly, note that $b_{\rm UV} = 1$ is mapped to  $b_{\rm IR} = 0$,  where $\Gamma_{\rm IR} = 2$,  thus the $b_{\rm UV} = 1$ theory is marginally irrelevant in the IR, consistent with a massless flow. 

One can also argue that the background charge must be $\Q = \alpha_1 (b+1/b -2)$ with $\alpha_1 =1$ as follows. 
We identify $\Gamma = 2 b (\Q -b)$ based on above considerations.   The RG flows predict $\Gamma_{\rm IR} = - \Gamma_{\rm UV}$,  which is a complicated relation between $b_{\rm IR}$ and $b_{\rm UV}$ for generic value of $\alpha_1$.    One can check that unless $\alpha_1 =1$, for $b_{UV}>1$, $b_{\rm IR}$ is generally complex.   Only for $\alpha_1 =1$ does one have the simple relation $b_{\rm IR} = 1- b_{\rm UV}$.

\subsection{Remarks on real verses imaginary $\mu$}

Let use make a few remarks concerning the $g_1$ real case originally considered in \cite{BL}.   
First of all,  the massless flows for the imaginary sine-Gordon theory no longer exist,  since for small $g_{1,2}$ the RG trajectories based on the RG invariant $\II$ are no longer approximately circles,  but rather hyperbolas.   
However the flows that begin at $g_1 = 0$ and end up at $g_1 = \infty$ have the same endpoints,  and the relations between
$\Gamma_{\rm UV}$ and $\Gamma_{\rm IR}$ presented above remain the same.   
This can be seen from the fact that equations \eqref{Qlines} are the same.   However the detailed trajectories are different.   
One can easily see with contour plots of $\II$ for real verses imaginary $g_1$ that the topologies of the flows in Figures
\ref{sinhMassive} and \ref{sinhFig} are essentially interchanged.  

\section{Application to the freezing transition in disordered systems}
\label{disorder}

We now make contact with disordered systems and explain the relation between the above sinh-Gordon model, with imaginary coupling, and Dirac fermions in random gauge field. See also \cite{Doussal,Doussal2}. This was actually our motivation when we started looking at this problem fifteen years ago, and left  it aside for a short  while.

\subsection{Dirac fermions in a random U(1) gauge field}

We consider two-component Dirac fermions in a random gauge field $A_\mu$ in two spatial dimensions $(x,y)$ plus time.  Defining complex spatial coordinates $z=(x+i y)/\sqrt{2}$ and $\zbar = (x-i y)/\sqrt{2}$,  the model is defined by the 
random hermitian hamiltonian
\beq
\label{Ham}
H = 
\( 
\begin{matrix}
0 & - i \d_\zbar + A_\zbar \\
-i \d_z + A_z & 0 \\
\end{matrix}
\).
\eeq
The probability distribution will be specified below.

The Green functions,  Fourier transformed in time to energy $\CE$,  are given by functional integrals with respect to the action
\beq
\label{SHam}
\CS = i \int  \frac{d^2 x}{2 \pi} \Psi^\dagger \( H - \CE \) \Psi .
\eeq
Introducing component fields as follows, $\Psi = \(\begin{smallmatrix}\psi_+ \\ \psibar_+ \\ \end{smallmatrix} \)$ and $\Psi^\dagger = \( \psibar_-, \psi_- \)$, 
one finds 
\beq
\label{Action2}
\CS(\Psi,A) = \int \frac{d^2 x}{2 \pi} \[ \, \psibar_- (\d_z - i A_z ) \psibar_+ 
+ \psi_- ( \d_\zbar - i A_\zbar ) \psi_+ 
+ i \CE \, (\psibar_- \psi_+ + \psi_- \psibar_+ ) \] .
\eeq

Disorder averaged correlation functions $\langle \CO \rangle$ are then defined as functional integrals over $A$:
\beq
\label{disave}
\bar{ \langle \CO \rangle} = \int D A \, P[A] \, \langle \CO\rangle_A 
\eeq
where the probability distribution for $A$ is taken to be gaussian:
\beq
\label{PA}
P[A] = \exp \( - \inv{g} \int \frac{d^2 x}{2 \pi} \, A_z A_\zbar \).
\eeq
The coupling constant $g$ is a measure of the strength of the disorder.   
In \eqref{disave}, $\langle \CO \rangle_A$ is the correlation function in a given realization of the disorder~:
\beq
\label{COA}
\langle \CO \rangle_A = \inv{Z(A, \CE) }\int D \Psi \, e^{-S(\Psi, A ) } \, \CO 
\eeq
where 
$Z(A, \CE)$ is the partition function.  

\subsection{Map to the sinh-Gordon model}

It is convenient to parameterize the gauge field in terms of a scalar field $\eta$ as follows\footnote{In $\mathbb{R}^2$, any gauge potential can be decomposed as $A_\mu = \d_\mu\vartheta +\half \epsilon_{\mu \nu} \d_\nu \eta$. But the pure gauge part $A_\mu = \d_\mu\vartheta$ can be gauged away in \eqref{Action2} and only the component $A_\mu = \half \epsilon_{\mu \nu} \d_\nu \eta$ matters.}~:
\beq
\label{Aeta}
A_\mu = \half \epsilon_{\mu \nu} \d_\nu \eta, ~~~~~
\Longrightarrow ~~~ P[\eta] = \exp \( - \inv{4g} \int \frac{d^2 x }{4 \pi} \( \d_\mu \eta \)^2 \)  .
\eeq
The coupling of the fermions to the gauge field can then be removed by the chiral gauge transformation:
\beq
\label{gaugetrans}
\psibar'_+ = e^{\eta/2} \, \psibar_+ , ~~~ \psibar'_- = e^{- \eta/2} \, \psibar_- , ~~~
\psi'_- = e^{\eta/2} \, \psi_- , ~~~ \psi'_+ = e^{-\eta/2} \, \psi_+ ~,
\eeq
and the action becomes 
\beq
\label{Action3}
\CS = \int \frac{d^2 x}{2 \pi} \[  \psibar'_- \d_z \psibar'_+ + \psi'_- \d_\zbar \psi'_+ + i \CE 
\( e^\eta \, \psibar'_- \psi'_+  + e^{-\eta} \, \psi'_- \psibar'_+ \) \]  .
\eeq

To make further progress,  we first consider $\CE$ to be very small,  and later restore it as a perturbation. 
When $\CE =0$,  the jacobian which arises in passing from $\Psi$ to $\Psi'$ in the functional integral precisely cancels the
$1/Z$ factor in \eqref{COA}.   This is easily seen by bosonizing the fermions $\Psi$ with a single boson $\phi$ so that the action 
\eqref{Action2} becomes (when $\CE =0$):
\beq
\label{Actionphieta}
\CS = \int \frac{d^2 x}{4 \pi} \Big( \inv{2} \( \d \phi \)^2 + i \d \eta \d \phi \Big) .
\eeq
The functional integrals over $\Psi$ and $\Psi'$ are then simply related by the shift $\phi \to \phi'-i \eta$.  

When $\CE = 0$ the functional integrals over $\Psi'$ can be done and do not introduce any new $\eta$ dependence. 
To restore the $\CE$ perturbation,  we make a mean field approximation and replace the $\Psi'$ fermion bilinears by
their one-point functions in a finite geometry of size $L$.   Since the fermions have dimension $1/2$, we have ~:
\beq
\label{onepoint}
\langle \psibar'_- \psi'_+ \rangle \sim \langle \psi'_- \psibar'_+ \rangle \sim 1/L ~.
\eeq
One is finally left with the functional integral over $\eta$.   Rescaling $\eta = \sqrt{2g} \, \phi $, one finds the sinh-Gordon action 
\beq
\label{SsinhG}
\CS [ \phi ] = \int d^2 x \( \inv{8 \pi} \( \d \phi \)^2 + 2 \mu \cosh \big( \sqrt{2}\,  b\, \phi \big) \) ~,
\eeq
where
\beq
\label{shGparams}
b = \sqrt{g} , ~~~~ \mu = i \, \frac{\CE}{2\pi  L} ~.
\eeq
The density of states operator is the one that couples to $\CE$,  which we chose to normalize as follows~:
\beq
\label{rhodef}
\rho \equiv  \frac{1}{L} \cosh \big( \sqrt{2} \, b \, \phi \big) ~.
\eeq
By definition one has $\bar{ \langle \rho \rangle } = \int D \phi e^{-S[\phi]} \, \rho$.

\subsection{Multi-fractal density of states exponents}

We first review some general standard definitions of exponents characterizing the density of states.  
Let $\rho (x)$ denote the density of states field operator.   The physical density of states is its vacuum expectation value,
i.e. the 1-point function denoted as $\langle \rho \rangle$, and depends on the realization of the disorder.   
Let $\bar{ \langle \rho \rangle }$ denote the disorder averaged quantity.  For a system of size $L$, one defines the
fundamental exponent $\Gamma_1$ as 
\beq
\label{Gamma1}
\bar{\langle \rho \rangle } \sim L^{-\Gamma_1} .
\eeq
In other words,  the exponent $\Gamma_1$ is just the anomalous dimension of the operator $\rho$ in the disorder averaged theory.   

Also of interest are multi-fractal exponents $\Gamma_q$ defined as follows~:   $\Gamma_q$ is defined as the anomalous dimension of the $q$-th moment of $\rho$:
\beq
\label{Gammaq}
\Gamma_q = \dim{\, \bar{\langle \rho \rangle^q}\,} ~,
\eeq
where we use the same notation as above, where $\dim{X}$ denotes the scaling dimension of $X$ in inverse length units. 

Because it is related to the multi-fractal spectrum of the density $\langle \rho (x) \rangle$, or of the associated measure $\langle \rho (x) \rangle\, dx$, a related quantity that is often studied is the normalized ratio 
\beq
\label{Pq} 
P^{(q)} =  \frac{ \int d^2 x \, \bar{ \langle \rho (x) \rangle^q}}
{\( \int d^2 x \, \bar{\langle \rho (x) \rangle } \)^q} ~.
\eeq
Simple scaling leads to 
\beq
\label{PqTau}
P^{(q)} \sim L^{- \tau (q) }
\eeq
where
\beq
\label{tau}
\tau (q) = \Gamma_q - q \Gamma_1 + 2 (q-1).
\eeq
Legendre transform of $\tau(q)$ gives access to the spectrum of multi-fractal dimensions of the density $\langle \rho (x) \rangle$.

\subsection{Multi-fractal spectrum}

Returning to our model of interest,  using the mapping to the sinh-Gordon model and \eqref{rhodef}, we have
\beq
\label{multi1}
\Gamma_1 (g) = 1 + \gamma (g) ~,
\eeq
where the ``$1$" comes from the $1/L$ in \eqref{rhodef}, and $\gamma (g)$ the scaling dimension of $\cosh ( \sqrt{2g} \, \phi )$, 
\beq
\label{gammadef}
\gamma (g) = \dim{ \cosh ( \sqrt{2g} \, \phi ) } ~.
\eeq
For higher $q$,  since the leading term in $\rho^q$ is $\cosh ( q \sqrt{2g} \, \phi )/L^q$, one has
\beq
\label{multi2}
\Gamma_q (g) = q + \dim{ \cosh (q \sqrt{2g} \, \phi )} ~.
\eeq
Since in the above equation the $\cosh$-operator is related to $\rho$ by $g \to q^2 g$,  this immediately leads to the fundamental  equation 
\beq
\label{multi3}
\Gamma_q (g) = \Gamma_1 (q^2 g) + q-1 .
\eeq
The latter implies 
\beq
\label{tau2}
\tau (q) = \Gamma_1 (q^2 g ) - q \Gamma_1 (g) + 3(q-1) .
\eeq
Given \eqref{multi1},  one sees that everything boils down to the dimension of the $\cosh(\sqrt{2} \, b \phi )$ operator in
the sinh-Gordon theory.  According to our proposal for the freezing transition in sinh-Gordon, we have $\Gamma_1 (g) = 1 -2g$ for $g<1$ and $\Gamma_1 (g) =3-4\sqrt{g}$ for $g>1$.

Transitions in the variable $q$ are thereby related to transitions in $b=\sqrt{g}$.    Using our proposal for a freezing transition in
the sinh-Gordon model 
\eqref{dimcases},  the 
two transition points are then $b=g=1$ and $q^2 g = q^2 b^2 =1$.     There are thus $4$ distinct regimes.   In terms of 
$g$ and $q$, they are:
\begin{subequations}
\begin{align}
&~ g<1, ~~ q< 1/\sqrt{g}: ~~~~~~~~~~ \tau(q) = 2 (q-1)(1-qg) \\
&~ g<1, ~~ q> 1/\sqrt{g}: ~~~~~~~~~~ \tau(q) = 2 q (1-\sqrt{g})^2  \\
&~ g>1, ~~ q< 1/\sqrt{g}: ~~~~~~~~~~ \tau(q) = -2 (1-q\sqrt{g})^2  \\
&~ g>1, ~~ q>1/\sqrt{g}: ~~~~~~~~~~ \tau(q) = 0.
\end{align}
\end{subequations}
This agrees with known results  \cite{Doussal,Doussal2,Castillo}.  

\section{Summary and discussion}

We have presented a specific proposal for the behavior of the sinh-Gordon model above the self-dual point $b>1$ that is quite different from the analytic continuation $b\to 1/b$ of the well-understood properties of the massive theory 
for $0<b<1$.     The main properties of this theory is that unlike the $0<b<1$ region it has a non-zero  background charge $\Q$ given in \eqref{Qinf}.    The theory is massless but not conformally invariant,  but rather is a relevant perturbation in the UV that flows
to another CFT in the IR,  arriving there via an irrelevant operator.     We provided two supporting arguments. 
The first was based on the beta functions in \cite{Moriconi,BL},   which are ultimately based on perturbation theory for the
sinh-Gordon action,  and do not show a $b \to 1/b$ symmetry,  and clearly predict different RG flows for $b<1$ verses $b>1$.
The second is that our proposal correctly reproduces known exact results for a Dirac fermion in a random magnetic field,
in particular all the transitions in the multi-fractal exponents.

 If our proposal is indeed correct,  it remains to determine the S-matrices for the massless flow when $b>1$ along the lines
 formulated in \cite{ZZMassless}.     This is beyond the original scope of this paper,  however there are some natural guesses. 
Letting $L$ and $R$ signify left verses right movers as in \cite{ZZMassless},  it is likely that the LL and RR S-matrices are 
$S_{\rm LL} = S_{\rm RR} = S_{\rm shG}$ where $S_{\rm shG}$ is the function of rapidity in \eqref{Smatrix}.        
This would guarantee that in the IR,  $c=1$.     It remains to specify left-right scattering $S_{\rm LR}$ which controls the UV. 
   It is natural to consider
$S_{\rm LR} = S_{\rm shG}$ here also,  however there are clearly other possibilities to be explored,  such as 
the very simplest possibility $S_{\rm LR} (\theta ) = - \tanh \( \tfrac{\theta}{2} -  \tfrac{i \pi}{4} \)$.    Clearly more work needs to be done in this direction.   

\bigskip

There are some natural questions that would be worthwhile to investigate to provide further support for our proposal.  
 We can think of these:
 
 \bigskip
 \n $\bullet$ 
 We should say  that the validity of the  beta functions we used in Section \ref{AllOrdersBeta} and our interpretation of 
 scaling dimensions at 
 $g_1 = \infty$,  namely based on \eqref{GammaIR},  could benefit from closer scrutiny,  even though we showed how these beta functions can reproduce known exact results on massless flows in the sine-Gordon model \cite{FSZ1,FSZ2}.   
 We refer to the Introduction for further remarks about this.

 \bigskip
 \n $\bullet$.  
     Konik et. al. \cite{KonikMussardo} essentially showed that  for the sinh-Gordon theory,   perturbation theory of the Liouville 
 theory and the free gaussian field agree in the weak coupling region $b<1$.   Can this analysis be extended to $b>1$ with the different background charge proposed here?
 
 \medskip
 
 \n $\bullet$    Can the semi-classical analysis in the Appendix be extended to higher order in perturbation theory?   It's unlikely this can fully confirm our exact proposal to all orders,  but a few low orders could provide convincing evidence.  

\medskip
\n  $\bullet$    Our suggestion in the last paragraph for the exact S-matrix clearly needs more investigation.   A clear way to proceed is with the Thermodynamic Bethe Ansatz.  

\medskip
   
 \n $\bullet$   
  It would be interesting to 
 investigate the problem by completely different  means, for instance from a lattice formulation of the sinh-Gordon model, or using continuous network tensor techniques adapted to field theory \cite{CiracVerstraete,Tilloy}.   Or,   perhaps a   rigorous probabilistic construction as in \cite{LiouvilleChaos} is possible.   
 
 \bigskip

 There are other possible applications of the freezing transition that our work may shed some light on. 
 An obvious one is to more complicated disordered systems such as the quantum Hall transition.   
 We also mention that it has been applied  to extreme values of the Riemann zeta function \cite{Keating1,Keating2}.  
 
\section{Acknowledgements}

We would like to thank Pierre Le Doussal,  Giuseppe Mussardo, Henri Orland, Kostas Sfetos, Kostas Siampos, Germ\'an Sierra and  Alyosha Zamolodchikov for discussions.  Alyosha first suggested to us  long ago the check of our beta functions based on comparison with massless flows in the imaginary sine-Gordon model.   
AL is thankful for the support of the Ecole Normale Sup\'erieure  in Paris and the organizers of the celebration of DB's 60-th birthday which led to a re-examination of this work we started  and abandoned 15 years ago;  he  also wishes to thank the Scuola Internazionale Superiore di Studi Avanzati (SISSA)  in Trieste, Italy, for support while this work was completed.

\appendix

\section{Semi-classical freezing and the Manning condensation}

Let us imagine computing semi-classically the one-point function of an exponential operator in the sinh-Gordon theory. Restoring  $\hbar$ so that the action becomes $\CS\to {\hbar}^{-1}\CS$, the one-point function of the operator $\exp({\sqrt{2}a}{\hbar}^{-1}\phi)$, $a>0$, located at the position $x_0$, is represented by the functional integral
\beq
\int D\phi\, e^{-{\hbar}^{-1}\big(\CS- \sqrt{2}a\phi(x_0)\big)} ~.
\eeq
In the semi-classical limit $\hbar\to0$, the integral is dominated by the saddle point field configuration $\phi_\mathrm{cl}$, solutions of
\beq \label{eq:classical}
-\frac{1}{4\pi} \Delta_x\phi_\mathrm{cl}(x) + 2\hat \mu\, \sinh(\sqrt{2}b\phi_\mathrm{cl}(x)) = \sqrt{2}a\, \delta^{(2)}(x-x_0) ~,
\eeq
where $\Delta_x$ is the Laplacian in 2D and $\delta^{(2)}(x-x_0)$ the Dirac measure at $x_0$ and $\hat \mu = \sqrt{2}b\mu$. Equation \eqref{eq:classical} can be solved exactly using tau function techniques \cite{TracyWidom}, but we do not need this explicit solution for the simple argument we now present. To take care of the $\delta$-function source, we should have $\phi_\mathrm{cl}(x)\simeq -\sqrt{2}a\log|x-x_0|^2$ as $|x|\to x_0$. Thus we set $\phi_\mathrm{cl}(x)= -\sqrt{2}a\log|x-x_0|^2 + \varphi(x)$, with $\varphi(x)$ sub-leading near $x_0$. We take $\varphi$ decreasing as a power law, so that
\beq
\phi_\mathrm{cl}(x)= -\sqrt{2}a\log|x-x_0|^2 + c_0 + c_1\, |x-x_0|^\sigma + \cdots ~,
\eeq
with $c_0$, $c_1$ two constants and $\sigma>0$ (so that $\varphi$ is sub-leading as $x$ approaches $x_0$) and where the dots refer to higher sub-leading terms near $x_0$.
The exponent $\sigma$ is found by matching the leading terms in $\Delta_x\phi_\mathrm{cl}$ and in $\sinh(\sqrt{2}b\phi_\mathrm{cl})$. This yields
\beq
|x-x_0|^{\sigma-2} \sim e^{-2ab\log|x-x_0|^2}= |x-x_0|^{-4ab}\quad \Longrightarrow ~ \sigma= 2(1-2ab) ~.
\eeq
Since we should have $\sigma>0$, this is possible only for $a<a_c=1/2b$. For $a>a_c$, the operator $\exp({\sqrt{2}a}{\hbar}^{-1}\phi)$ is actually screened such that its effective weight $a_\mathrm{eff}$ at large scale is $a_c$. 

This semi-classical computation indicates the possibility of a freezing transition. For any fixed sinh-Gordon parameter $b$, the exponential operators $\exp({\sqrt{2}a}\,\phi)$ are well-defined for $a<a_c$ only, for some critical value $a_c$, but they get frozen for $a>a_c$ to the critical exponential operator $\exp({\sqrt{2}a_c}\,\phi)$ with critical weight $a_c$. In view of the symmetry relation \eqref{Reflection}, valid in Liouville theory, it is tempting to propose that $a_c=\Q/2$. This is compatible with the semi-classical limit $a_c\simeq 1/2b$ for $b\to 0$. 

This phenomena is known in the physics of polyelectrolyte solutions as the Manning condensation \cite{Manning}. Imagine considering a positively charged polymer, say a DNA, immersed in a polyelectrolyte made of positive and negative charged ions, and ask what is the electrostatic potential for this system. If we imagine the polymer to be straight along the $z$-axis, then \eqref{eq:classical} is the Poisson-Boltzmann equation for this electrostatic problem in the 2D transverse directions.  If the charge density of the polymer is too high,  larger than a critical value $a_c$, it is screened by oppositely  charged ions which occupy a cylindrical volume around the polymer of diameter $r_c$, so that the system formed by the polymer and these counter-ions behaves at a distance higher than $r_c$ like a polymer of critical charge density $a_c$. This is the Manning's screening effect.


\begin{thebibliography}{99}



\bibitem{sinhG1}
A. E. Arinschtein, V. A. Fateev and A. B. Zamolodchikov, 
{\it Quantum S-matrix of the (1+1) Dimensional Toda Chain} 
Phys. Lett.   {\bf 87B} (1979)  389.  

\bibitem{MussardoSinh1}
A. Koubek and G. Mussardo, 
{\it On the Operator Content of the Sinh-Gordon model},
Phys. Lett. B {\bf 311} (1993), 193
[arXiv:hep-th/9306044].

\bibitem{MussardoSinh2}
A. Fring, G. Mussardo and P. Simonetti,
{\it  Form factors for integrable lagrangian field theories, the sinh-Gordon model},
Nucl. Phys.  {\bf B393} (1993) 
[arXiv:hep-th/9211053]. 

\bibitem{LeClairMussardo} 
A. LeClair and G. Mussardo, 
{\it Finite temperature correlation functions in Integrable QFT}, 
Nucl. Phys. {\bf B552} (1999) 624.
[arXiv:hep-th/9902075].    

\bibitem{ZamoTBAsinhG}
Al. B. Zamolodchikov,
{\it On the thermodynamic Bethe ansatz equation in the sinh-Gordon model,}
J. Phys. A: Math. Gen {\bf 39} (2006)  12863
[arXiv:hep-th/0005181].  

\bibitem{MussardoBook}
G. Mussardo,
{\it Statistical Field Theory,  An Introduction to Exactly Solved Models in Statistical Physics},
2010, Oxford University Press.   

\bibitem{KonikMussardo}
R. Konik, M. L\'ajer and G. Mussardo, 
{\it Approaching the Self-Dual point of the Sinh-Gordon model,}
[arXiv:2007.00154].  

\bibitem{Derrida}
B. Derrida, 
{\it Random-Energy Model: Limit of a Family of Disordered Models}, 
Phys. Rev. Lett. {\bf 45} (1980) 79;\\
B. Derrida,
{\it The random energy model, an exactly solvable model of disordered systems}, 
Phys. Rev. {\bf B24} (1981) 2613.

\bibitem{BouchaudFyodorov} 
Y Fyodorov and J.-P. Bouchaud, 
{\it Freezing and extreme value statistics in a Random Energy Model with logarithmically correlated potential}, 
J. Phys.A: Math. Theor {\bf 41} (2008) 372001.

\bibitem{Doussal} 
D. Carpentier and P. Le Doussal, 
{\it Glass transition of a particle in a random potential,  
front selection in non-linear RG and entropic phenomena in Liouville and sinh-Gordon models,}
Phys. Rev. {\bf E63} (2001) 026110
[arXiv:cond-mat/0003281]. 

\bibitem{Doussal2}
B. Horovitz and P. Le Doussal,
{\it Freezing transitions. and the density of states of 2D random Dirac hamiltonians,}
Phys. Rev. {\bf B65} (2002) 125323
[arXiv:cond-mat/0108143]. 

\bibitem{Moriconi}  B. Gerganov, A. LeClair and M. Moriconi, 
{\it On the beta function for anisotropic current interactions in 2D}, 
Phys. Rev. Lett. {\bf 86} (2001) 4753 [arXiv:hep-th/0011189]. 

\bibitem{BL}  D. Bernard and A. LeClair,  
{\it Strong-weak coupling duality in anisotropic current interactions,}
Phys.  Lett.{\bf  B512}  (2001) 78
[arXiv:hep-th/0103096]. 

\bibitem{FSZ1}
P. Fendley,  H. Saleur and Al. B. Zamolodchikov, 
{\it Massless Flows I: the sine-Gordon and O(n) models}, 
Int. J. Mod. Phys. {\bf A8} (1993) 5717
[arXiv:hep-th/9304050]. 

\bibitem{FSZ2}
P. Fendley,  H. Saleur and Al. B. Zamolodchikov, 
{\it Massless Flows II: the exact S-matrix approach}, 
Int. J. Mod. Phys. {\bf A8} (1993) 5751
[arXiv:hep-th/9304051]. 


\bibitem{Wiese} 
A. W. W. Ludwig and K. J. Wiese, 
{\it The 4-loop beta-function in the 2D Non-Abelian Thirring model, and comparison with its conjectured "exact" form,}
Nucl. Phys.  {\bf B661} (2003) 577 
[arXiv:cond-mat/0211531]. 

\bibitem{Chiral}
A. LeClair,
{\it Chiral stabilization of the renormalization group flow for flavor and color anisotropic current interactions,}
Phys. Lett. {\bf B519} (2001) 183,
[arXiv:hep-th/0105092].

\bibitem{Greeks1}
G. Itsios, K. Sfetsos and K. Siampos, 
{\it The all-loop non-Abelian Thirring model and its RG flow},
Phys. Lett. {\bf B733} (2014) 265
[arXiv:1404.3748].

\bibitem{Greeks2}
G.  Georgiou,  K. Sfetsos and K.  Siampos,
{\it  All-loop anomalous dimensions in integrable $\lambda$-deformed  $\sigma$-models},
Nucl. Phys. {\bf B901} (2015) 40
[arXiv:1509.02946].

\bibitem{Greeks3}
G. Georgiou, E. Sagkrioti, K Sfetsos and K. Siampos,
{\it An exact symmetry in $\lambda$-deformed CFTs},
DOI:10.1007/JHEP01(2020)083, 
[arXiv:1911.02027]. 

\bibitem{Tseytlin}
B. Hoare, N. Levine and A. Tseytlin,
{\it Integrable sigma models and 2-loop RG flow}
DOI: 10.1007/JHEP12(2019)146,
[arXiv:1910.00397] 

\bibitem{Castillo} 
H.  E. Castillo,  C. Chamon, E.  Fradkin, P.  M. Goldbart, and C.  Mudry,
{\it Exact calculation of multifractal exponents of the critical wave function of Dirac fermions in a random magnetic field,}
Phys. Rev. {\bf B56} (1997) 10668 
[arXiv:cond-mat/9706084]. 

\bibitem{Manning} 
G.S. Manning, 
{\it Limiting Laws and Counterion Condensation in Polyelectrolyte Solutions I. Colligative Properties}, 
J. Chem. Phys. {\bf 51} (1969) 924?933.

\bibitem{ZamoMassScale}
Al. B. Zamolodchikov,
{\it Mass scale in the sine-Gordon model and its reductions,}
Int. J. Mod. Phys. {\bf A10} (1995) 1125. 

\bibitem{LiouvilleShG1}
V.  Fateev,  S. L. Lukyanov, A. B. Zamolodchikov and Al. B. Zamolodchikov, 
{\it Expectation values of local fields in Bullough-Dodd model and integrable perturbed 
conformal field theories},
Nucl. Phys. {\bf B516} (1998) 652 
[arXiv:hep-th/9709034].  

\bibitem{LiouvilleShG2}
A. B. Zamolodchikov and A. B. Zamolodchikov, 
{\it Structure constants and conformal bootstrap in Liouville field theory,}
Nucl. Phys. {\bf B477} (1996) 577 
[arXiv:hep-th/9506136].

\bibitem{LiouvilleShG3}
G.  Mussardo and P.  Simonetti, 
{\it Stress-energy tensor and ultraviolet behavior in massive integrable quantum field theories},
Int. J.  Mod. Phys. {\bf A9} (1994) 3307 
[arXiv:hep-th/9308057]. 

\bibitem{ZZMassless}
A. B. Zamolodchikov and Al. B. Zamolodchikov,  
{\it Massless factorized scattering and sigma models with topological terms,}
Nucl. Phys. {\bf B379} (1992) 602. 

\bibitem{CiracVerstraete}
F. Verstraete and I. Cirac,
{\it Continuous Matrix Product States for Quantum Fields}, 
Phys. Rev. Lett. {\bf 104} (2010), 190405

\bibitem{Tilloy}
A. Tilloy, 
{\it Relativistic continuous matrix product states for quantum fields without cutoff}, 
Phys. Rev. {\bf D 104} (2021), 096007

\bibitem{LiouvilleChaos}
F. David, A. Kupiainen, R. Rhodes and V. Vargas, 
{\it Liouville Quantum Gravity on the Riemann sphere},
Commun. Math. Phys. {\bf 342} (2016) 869-907.

\bibitem{Keating1}
Y. V. Fyodorov,  G. A. Hiary and J. P. Keating,
{\it Freezing Transition, Characteristic Polynomials of Random Matrices, and the Riemann Zeta function,}
Phys. Rev. Lett. {\bf 108}, 170601 (2012),
[arXiv:1202.4713 [math-ph]. 

\bibitem{Keating2}
Y. V. Fyodorov and J. P.  Keating, 
{\it Freezing Transitions and Extreme Values:  Random Matrix Theory,  $\zeta (1/2 + it)$, and Disordered Landscapes,}
Phil. Trans. R. Soc. {\bf A372} (2014), 20120503,
[arXiv:1211.6063]. 

\bibitem{TracyWidom}
C. A. Tracy and H. Widom, 
{\it On exact solutions to the cylindrical Poisson-Boltzmann equation with applications to polyelectrolytes}, 
Physica {\bf 244A} (1997), 402-413.

\end{thebibliography}
\end{document}